\documentclass[prb,preprint,showpacs,amsmath,amssymb,superscriptaddress]{revtex4}
\usepackage{epsfig}
\usepackage{graphicx}
\usepackage{color}
\usepackage{amsmath,amssymb}

\newcommand{\beq}[1]{
\begin{equation}
\label{e#1} }

\newcommand{\eeq}{
\end{equation}
}
\newcommand{\rr}{{\bf r}}
\newcommand{\kk}{{\bf k}}
\begin{document}

\title{Spin Hall transistor with electrical spin injection}
\author{K.~Olejn\'{\i}k}
\affiliation{Hitachi Cambridge Laboratory, Cambridge CB3 0HE, United Kingdom}
\affiliation{Institute of Physics ASCR, v.v.i., Cukrovarnick\'a 10, 162 53
Praha 6, Czech Republic}

\author{J.~Wunderlich}
\affiliation{Institute of Physics ASCR, v.v.i., Cukrovarnick\'a 10, 162 53
Praha 6, Czech Republic}
\affiliation{Hitachi Cambridge Laboratory, Cambridge CB3 0HE, United Kingdom}

\author{A.~C.~Irvine}
\affiliation{Microelectronics Research Centre, Cavendish Laboratory, University of Cambridge, CB3 0HE, United Kingdom}

\author{R.~P.~Campion}
\affiliation{School of Physics and
Astronomy, University of Nottingham, Nottingham NG7 2RD, United Kingdom}

\author{V.~P.~Amin}
\affiliation{Department of Physics, Texas A\&M University, College
Station, TX 77843-4242, USA}

\author{Jairo~Sinova}
\affiliation{Department of Physics, Texas A\&M University, College
Station, TX 77843-4242, USA}
\affiliation{Institute of Physics ASCR, v.v.i., Cukrovarnick\'a 10, 162 53 Praha
6, Czech Republic}

\author{T.~Jungwirth}
\affiliation{Institute of Physics ASCR, v.v.i., Cukrovarnick\'a 10, 162 53
Praha 6, Czech Republic} \affiliation{School of Physics and
Astronomy, University of Nottingham, Nottingham NG7 2RD, United Kingdom}

\date{\today}



\pacs{72.25.Dc, 72.25.Hg, 85.75.Nn}

\maketitle
{\bf 
The realization of a viable semiconductor transistor and information processing devices based on the electron spin has fueled intense basic research  of  three key elements: injection, detection, and manipulation of spins in the semiconductor channel. The inverse spin Hall effect (iSHE) detection of spins injected optically in a 2D GaAs\cite{Wunderlich:2008_a,Wunderlich:2010_a}  and manipulated by a gate-voltage dependent internal spin-orbit field  has recently led to the experimental demonstration of a spin transistor  logic device.\cite{Wunderlich:2010_a} The aim of the work presented here is to demonstrate in one device the iSHE detection combined with an electrical spin injection and manipulation.  We use a 3D GaAs channel for which efficient electrical spin injection from Fe Schottky contacts has been demonstrated in previous works.\cite{Lou:2007_a,Ciorga:2008_a,Awo-Affouda:2009_a,Chan:2009_a,Salis:2009_a,Salis:2010_a,Garlid:2010_a} In order to experimentally separate the strong ordinary Hall effect signal from the iSHE in the semiconductor channel we developed epitaxial ultrathin-Fe/GaAs contacts allowing for Hanle spin-precession measurements in applied in-plane  magnetic fields. Electrical injection and detection is combined in our transistor structure with electrically manipulated spin distribution and spin current which, unlike the previously utilized electrical manipulations of the spin-orbit field\cite{Wunderlich:2010_a} or ballistic spin transit time,\cite{Huang:2007_a} is well suited for the  diffusive 3D GaAs spin channel. The magnitudes and external field dependencies of the measured signals are quantitatively analyzed using simultaneous spin detection by the non-local spin valve effect\cite{Lou:2007_a,Ciorga:2008_a,Awo-Affouda:2009_a,Chan:2009_a,Salis:2009_a,Salis:2010_a,Garlid:2010_a} and modeled by solving the drift-diffusion \cite{Yu:2002_a,Lou:2007_a} and Hall-cross response equations for the parameters of the studied microstructure.} 

The pioneering works  on electrical spin injection and detection\cite{Johnson:1985_a,Jedema:2001_a}  in non-magnetic channels were done in metals, taking advantage of the compatibility with conventional metal ferromagnets which formed the injection and detection electrodes. 
The non-local spin valve effect, utilized in these seminal studies, measures the dependence of the electro-chemical potential at the detection ferromagnetic electrode on the relative orientation of the magnetization of the electrode and the spins in the non-magnetic metal underneath it. 

A fundamentally distinct method for the electrical detection of spin currents in non-magnetic conducting channels is based on the iSHE.  The approach was first demonstrated  in metal devices with out-of-plane magnetized ferromagnetic injection contacts and compared in the same microstructure with the non-local spin valve signal.\cite{Valenzuela:2006_a,Seki:2008_a} 
The iSHE detection does not utilize a reference ferromagnetic probe. Instead,  a transverse spin dependent voltage  of a Hall cross fabricated directly in the non-magnetic channel provides the measure of the out-of-plane spin polarization  of the propagating electrons.  The basic physics distinction between the two approaches is that the spin valve effect originates from the exchange-splitting of carrier bands in the ferromagnetic probe while the iSHE originates from the spin-orbit coupling in the non-magnetic conductor.\cite{Dyakonov:1971_a,Murakami:2003_a,Sinova:2004_a,Kato:2004_d,Wunderlich:2004_a}

Several recent works have applied the non-local spin valve method for detecting spins in the GaAs semiconductor.\cite{Lou:2007_a,Ciorga:2008_a,Awo-Affouda:2009_a,Chan:2009_a,Salis:2009_a,Salis:2010_a}  We have prepared specially designed Fe/GaAs microstructures which allow us to reproduce these previous non-local spin valve experiments and to simultaneously demonstrate in the same microdevice the detection of the spin current by the iSHE in the lateral GaAs microchannel. An important feature of our Fe/GaAs devices which enables the iSHE detection is the strong in-plane anisotropy of the ultrathin-film (2~nm) Fe electrodes. It allows us to apply sufficiently strong in-plane magnetic fields along the Fe hard-axis for performing Hanle spin precession experiments, without aligning the Fe magnetization with the external field. This is the suitable geometry for the iSHE detection because spins injected from the in-plane magnetized Fe electrode precess in GaAs out of the plane of the transport channel. Furthermore,  the iSHE and the ordinary (Lorentz force) Hall effect contributions can be experimentally separated in this set up which is essential for detecting the iSHE in semiconductors. Recall that only in the high carrier density metals, the ordinary Hall effect is relatively weak and can be neglected in the iSHE experiments in external magnetic fields.\cite{Valenzuela:2006_a} 

Another important feature of our device is that it allows to manipulate the distribution of the diffusive spin current by applying electrical bias across the transport channel. Our work  complements the previous realization of the GaAs spin transistor\cite{Wunderlich:2010_a} in several aspects. (i) While the device reported in Ref.~\onlinecite{Wunderlich:2010_a} acted as a spin analogue of a field-effect-transistor logic element, the present device is rather a spin amplifier.  (ii) Our device does not utilize optical spin injection but electrical injection from a ferromagnetic contact. (iii) We measure simultaneously the output spin current by the iSHE and the output spin polarization by the non-local spin valve effect which allows us to quantitatively analyze the detected spin signals.

Our Fe/n-GaAs heterostructure was grown epitaxially in a single molecular-beam-epitaxy chamber without breaking ultra high vacuum conditions during the whole growth process. The sample was deposited on an undoped GaAs substrate. The heterostructure comprises 250~nm of low Si-doped GaAs ($5\times10^{16}$~cm$^{-3}$), followed by 15~nm of GaAs with graded doping, and 15~nm of highly Si-doped GaAs ($5\times10^{18}$~cm$^{-3}$). The doping profile yields a narrow tunnel Schottky barrier between GaAs and Fe favourable for spin injection/detection.\cite{Lou:2007_a,Awo-Affouda:2009_a,Chan:2009_a,Salis:2009_a} The growth temperature of GaAs was 580$^\circ$C. The sample was then cooled to 0$^\circ$C for the growth of the 2~nm Fe layer. The reflection high energy electron diffraction pattern observed after the Fe deposition confirmed the epitaxial growth of cubic Fe. The Fe film was  capped by a 2~nm Al layer to prevent Fe oxidation. Electron-beam lithography and wet chemical and reactive ion etching were used to pattern the lateral GaAs channel with the Hall crosses and magnetic electrodes, shown schematically in Fig.~1a (for more details see Supplementary information). The distance between Fe electrodes is  4~$\mu$m and between the Fe electrode and the Hall cross the distance is 2~$\mu$m. Non-magnetic Au-contacts are located 100~$\mu$m away from the central Fe electrode on  each side of the channel. All measurements shown in this paper were performed at 4.2~K. The reproducibility of our experimental data was confirmed by performing measurements in three different samples with the same nominal  heterostructure parameters and microdevice geometry and in each sample by contacting different combinations of available electrodes. 

\begin{figure}[h!]
\vspace*{-2cm}
\hspace*{0cm}\epsfig{width=0.5\columnwidth,angle=0,file=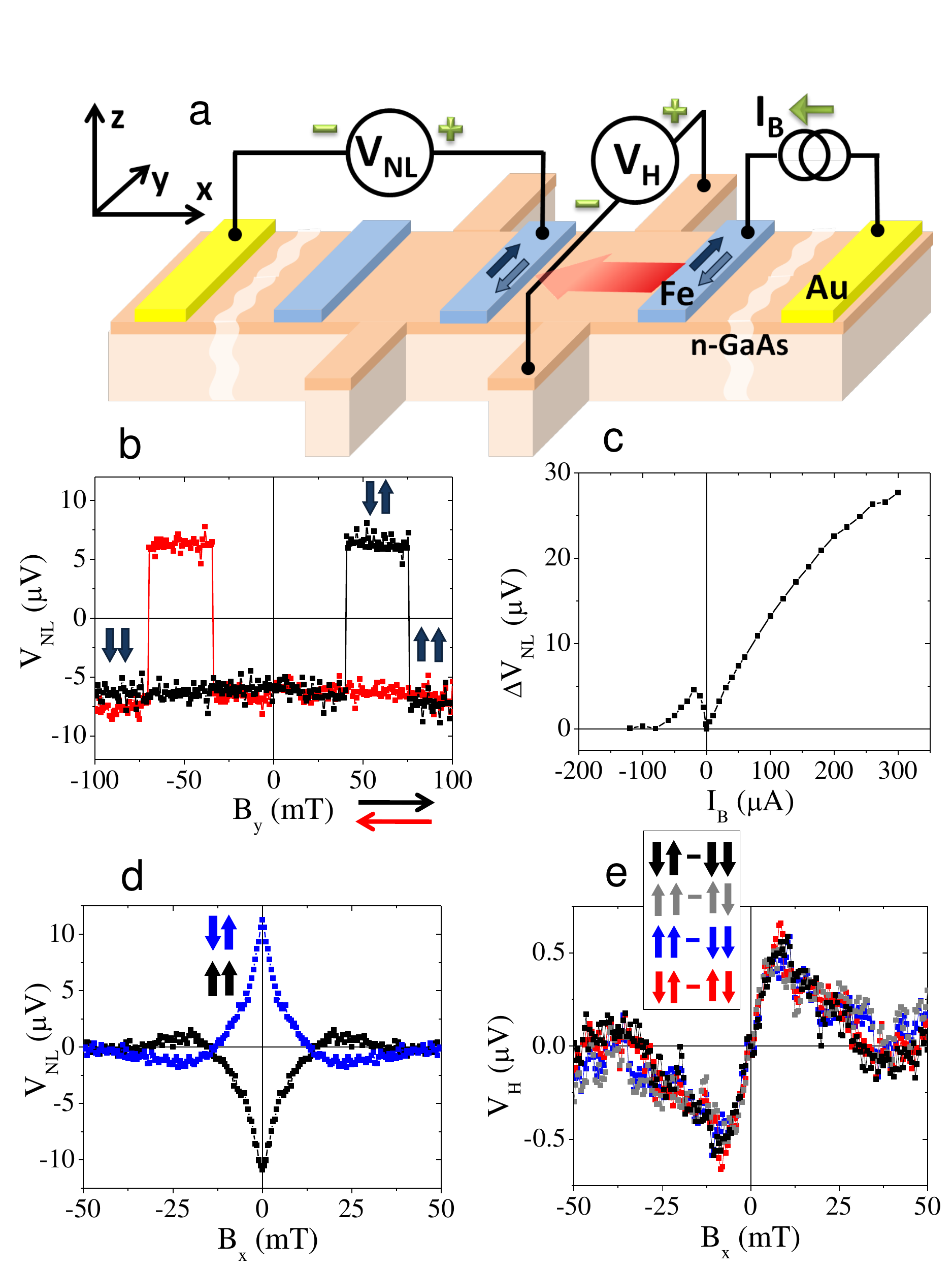}
\vspace*{0cm}
\caption{
(a) Schematics of the device for electrical spin-injection via a bias current $I_{B}$, and the simultaneous detection of the spin current by the iSHE voltage $V_{H}$ and the spin accumulation by the non-local spin valve voltage $V_{NL}$. Green symbols show the notation corresponding to positive voltages and current. (b) $V_{NL}$ measured by sweeping the external magnetic field $B_y$ along the Fe easy-axis. The left (right) black arrow in each pair of arrows inside the plot indicates the magnetization state of the detection (injection) Fe electrode. The black and red arrows next to the label of the $B_y$-axis indicate the field sweep direction. The data were measured at $I_B=100$~$\mu$A. (c) The difference between the non-local spin-valve signals at antiparallel and parallel configurations of magnetization in the injection and detection Fe electrode as a function of the spin-injection bias current.  (d) Hanle spin precession/dephasing measurement by the non-local spin valve voltage in an applied in-plane hard-axis field $B_x$. As in panel (b), the arrows inside the plot indicate the magnetization states of the detection and injection Fe electrodes preset before the Hanle measurements. The color coding of the arrow pairs corresponds to the color of the measured data.  The data were measured at $I_B=300$~$\mu$A and magnetic field sweep rate 2~mT/min. (e) Measured iSHE signal under same experimental conditions as in (d). The arrows notation is the same as in panel (d).
 }
\label{eb1}
\end{figure}

The magnetic anisotropy of our Fe electrodes has the strong out-of-plane component (2~T) due to the thin-film shape anisotropy, the cubic magnetocrystalline component, and an  additional  uniaxial magnetocrystalline anisotropy originating from the broken [1$\bar{1}$0]/[110] symmetry of our ultrathin Fe on GaAs. These anisotropies make the [110] in-plane crystal direction ($y$-axis in Fig.~1a) the easy magnetic axis with an anisotropy field of 0.2~T required to align the magnetization with the [1$\bar{1}$0] in-plane hard-axis ($x$-axis in Fig.~1a). This anisotropy field is significantly larger than external magnetic fields applied in the Hanle precession experiments with typical amplitudes up to 50~mT.  When applied along the in-plane hard-axis, the component of the magnetization along the easy-axis is reduced by less than 10\%. (For a more detailed discussion of the magnetic anisotropies see Supplementary information).

For characterizing individual Fe/GaAs Schottky contacts we performed three-point measurements between an individual Fe electrode and the two Au electrodes. These measurements confirm the tunneling nature of transport through the Schottky barrier. By measuring the tunneling anisotropic magnetoresistance of individual Fe/GaAs Schottky contacts,\cite{Moser:2006_a} we inferred the strength of the above anisotropy fields and the technique also allows us to infer the order in which the Fe electrodes switch their magnetization 
(see Supplementary information). 
The magnetic characterization of each individual Fe electrode is essential for the interpretation of the spin injection, manipulation, and detection measurements which we now discuss in detail. 

In Fig.~1b we plot the non-local spin-valve signal  measured while sweeping the external magnetic field $B_y$ along the easy axis of the Fe electrodes. In this geometry, the field triggers magnetization reversal in the Fe electrodes via a domain nucleation and propagation process. The reversal fields are different in different Fe electrodes which allows us to control independently the magnetization orientations in the injection and detection electrodes. In the experiment, a bias current $I_{B}$ driven between the injection (right) Fe/GaAs Schottky contact and the right Au electrode generates spin-accumulation underneath the spin-injection contact. A resulting diffusive spin current propagates into the unbiased part of the semiconductor channel with the Hall cross and the detection (central) Fe electrode. The lower value of the non-local voltage, measured between the detection Fe electrode and the left Au electrode, corresponds to parallel orientations of magnetization of the injection and detection Fe electrodes; the higher value of the non-local voltage corresponds to antiparallel magnetizations. These data together with the dependence of the amplitude of the non-local voltage on the bias current $I_{B}$, plotted in Fig.~1c, reproduce previous results of spin injection experiments in GaAs channels with Fe Schottky contacts.\cite{Lou:2007_a,Awo-Affouda:2009_a,Salis:2009_a,Garlid:2010_a} Note that at $I_{B} > 0$, which is more favorable for efficient spin injection, electrons in the biased part of the channel drift in the direction from GaAs to Fe, i.e., are spin-selectively extracted from the semiconductor. 

Non-local spin valve measurements in magnetic fields $B_x$ applied along the Fe in-plane hard axis  are shown in Fig.~1d. The black curve shown in the figure was obtained by setting the magnetizations in the Fe injection and detection electrodes in the parallel configuration before sweeping the in-plane hard-axis field; the blue curve was measured in the antiparallel magnetization configuration.  At zero field we obtain the higher value of the non-local voltage for antiparallel magnetizations, consistent with Fig.~1b. 

The Hanle dependence of the non-local spin valve signal on the hard-axis field $B_x$ shown in Fig.~1d reflects the precession and dephasing of spins in the GaAs channel, as quantified in detail in the theory section. The injected spins precess in the plane perpendicular to the applied field $B_x$, i.e., acquire an out-of-plane component. Our observation of the iSHE signal due to the out-of-plane polarized spin current in  the GaAs is demonstrated in Fig.~1e. Consistent with the iSHE interpretation, the signal in Fig.~1e is zero at zero applied field since in this case the in-plane polarized injected spins do not precess in the GaAs channel, i.e., do not acquire the out-of-plane component. The variations of the iSHE signal in Fig.~1e and of the Hanle non-local spin valve signal in Fig.~1d occur at a comparable magnetic field scale. This confirms the precession origin of the  out-of-plane spin component detected by the iSHE voltage. A full quantitative modeling of these Hanle non-local spin valve and iSHE curves is discussed in the theory section. 

The iSHE curves shown in Fig.~1e were obtained by subtracting the measured signals for oppositely preset  polarizations of the Fe injection electrode before performing the Hanle measurement. The individual raw data measured for a given orientation of the injector polarization have a linear contribution from the ordinary Hall effect. Importantly, the ordinary Hall contribution is independent of the preset orientation of the magnetization of the injection electrode and is therefore experimentally removed by subtracting the data for opposite injector polarizations. The signals shown in Fig.~1e are, therefore, of pure spin origin and are due to the iSHE which is an odd function of the polarization of injected electrons. (For the comparison of the measured raw data and the iSHE signal and the discussion of  the ordinary Hall effect in our sample geometry see Supplementary information).  Since spin detection by the iSHE is performed directly in the GaAs channel, the corresponding signal depends on the magnetization state of the Fe injection electrode and, unlike the non-local spin valve signals, it is independent of the state of the Fe detection electrode. This expected behavior of the iSHE data is confirmed in Fig.~1e which shows measurements for different magnetization configurations of the injection and detection electrodes. 

\begin{figure}[h]
\hspace*{0cm}\epsfig{width=0.5\columnwidth,angle=0,file=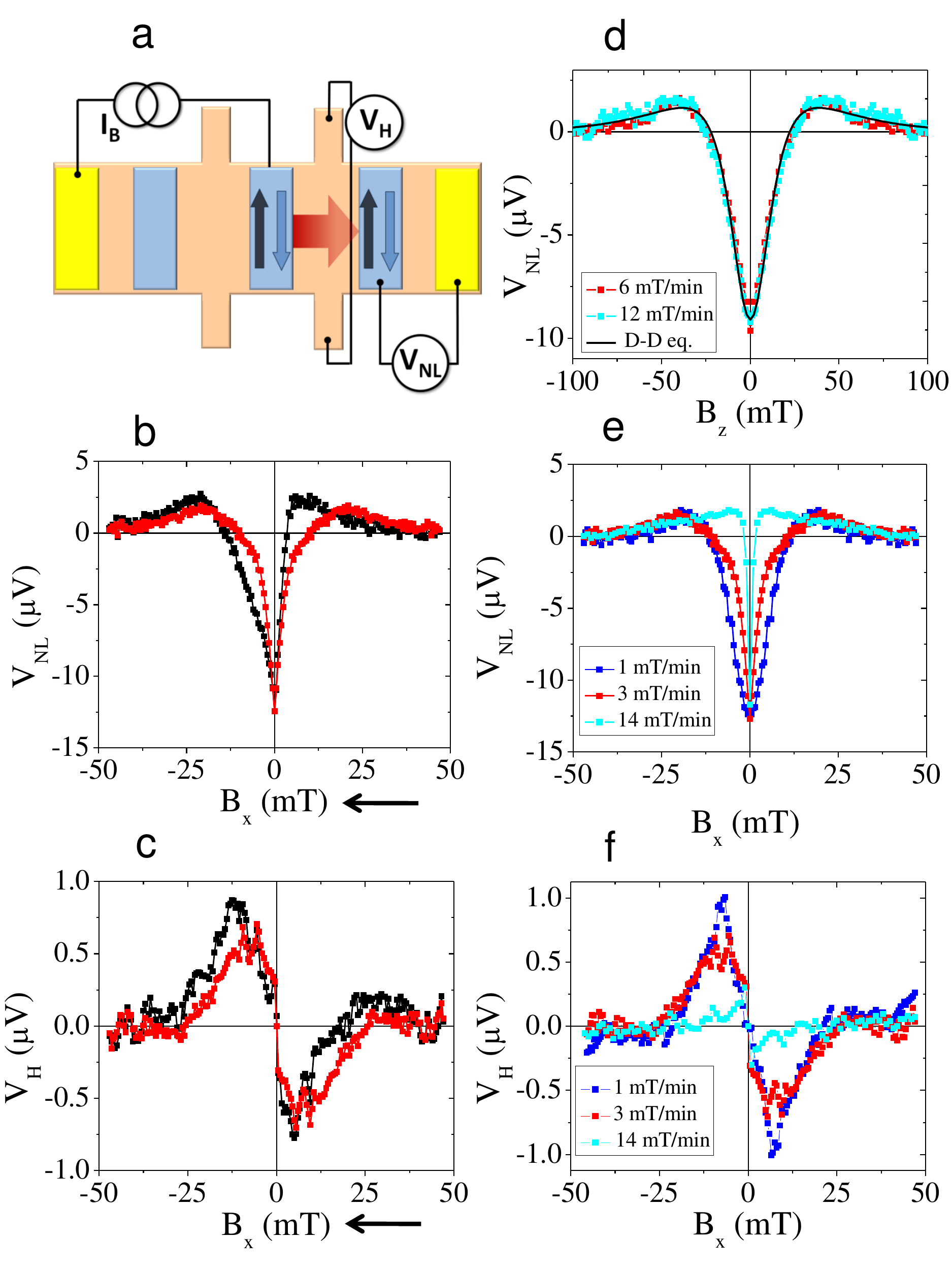}

\vspace*{0.5cm}
\caption{
(a) Schematics of the experimental setup. (b) As measured (black) and symmetrized (red) non-local spin valve signal in the Hanle experiment with the in-plane hard-axis field $B_x$. The black arrow next to the label of the $B_x$-axis indicates the field sweep direction. The data were measured at $I_B=300$~$\mu$A and magnetic field sweep rate 3~mT/min. (c) As measured (black) and antisymmetrized (red) iSHE signals for the same experimental conditions as in (b).  (d) Symmetrized non-local spin valve signal in the Hanle experiment with the out-of-plane hard-axis field $B_z$ measured at different sweep rates. The solid line shows the solution of the drift-diffusion equations discussed in the theory section. (e),(f) Symmetrized non-local spin valve measurements and antisymmetrized iSHE measurements in the in-plane hard-axis field $B_x$ at different field sweep rates.
 }

\label{eb2}
\end{figure}

For clarity, the Hanle non-local spin valve  data shown in Fig.~1d are symmetrized and, similarly, the iSHE data in  Fig.~1e are antisymmetrized with respected to $B_x=0$. In Figs.~2b,c we show that the symmetrization (antisymmetrization) does not affect the key features of the measurements (see also Supplementary information). Nevertheless, the absolute value of the raw data have a noticable asymmetry which we attribute to the presence of dynamic nuclear spin polarization effects.\cite{Awo-Affouda:2009_a,Chan:2009_a,Salis:2009_a}  To highlight the presence of the nuclear spins we plot  in Figs.~2d,e the Hanle curves measured at different magnetic field sweep rates. For the applied in-plane field $B_x$, the Hanle curves show strong dependence on the sweep rate, consistent with the presence of the nuclear spin effects. On the other hand, the Hanle curves measured in the out-of-plane  field $B_z$ are independent of the sweep rate, implying that the nuclear spins do not significantly contribute in this experimental geometry. 

The role of nuclear spins in the measured Hanle curves can be understood from the following expression for the Overhauser field ${B}_n$ they produce and which renormalizes the total effective field acting on electron spins,\cite{Salis:2009_a,Farah:1998_a}
\begin{equation}
{\bf B}_n=fb_n\left(\hat{\bf B}\cdot\langle{\bf S}\rangle\right)\hat{\bf B}\,.
\label{nuclear}
\end{equation}
Here $f\le1$ is a nuclear-spin relaxation leakage factor, $b_n=-8.5$~T in GaAs, $\hat {\bf B}$ is the unit vector of the external magnetic field, and $\langle{\bf S}\rangle$ is the mean electron spin polarization. Because of the large out-of-plane magnetic anisotropy of the thin-film Fe (with the corresponding 2~T anisotropy field), the applied field $B_z$ in the out-of-plane Hanle experiment is not strong enough to significantly tilt the magnetization in the injection electrode from the easy-axis ($\hat y$-direction). The projection of the injected electron spins to the applied field $B_z$ remains small in this experimental geometry which explains the absence of nuclear spin effects in Fig.~2d. For the in-plane Hanle experiment, the projection is given approximately by $\langle S \rangle B_x/B_A$, where $B_A=0.2$~T is the in-plane anisotropy field. By fitting the solution of the drift-diffusion equation for electron spins to the measured non-local spin valve signals, shown in Fig.~2d and discussed in detail in the theory section, we infer that $\langle S \rangle\approx 5$~\%. This combined with the experimentally determined sign of the injected electron spin polarization yields $B_n\approx B_x$, i.e., the total effective field experienced by the electron spins in GaAs is, $B_x^{eff}\approx 2B_x$. (The sign of the injected electron spin polarization, which is opposite to the magnetization in the Fe injection electrode, was obtained using the tilted magnetic field Hanle measurements\cite{Salis:2009_a} which are discussed in the Supplementary information.) Consistently, the field scale on which the slow sweep rate Hanle curve varies in Fig.~2e is about a factor of 2 smaller than the field scale in Fig.~2d. 

Figs.~2e,f show the correspondence between the non-local spin valve and iSHE signals measured at different sweep rates of $B_x$. The correspondence demonstrates that the Overhauser field acts in both measurements which provides further evidence that the measured Hall signals are of spin origin. We conclude the discussion of nuclear spins by pointing out that at sufficiently low sweep rates the role of the nuclear spins is merely in rescaling the effective magnetic field acting on electron spins in GaAs. Apart from the rescaling, the nuclear spins do not obscure the results of our spin transport experiments. This applies to both the non-local spin valve and the iSHE detection measurements.  

We now inspect the key symmetry of the iSHE signal which is the change of the sign of the Hall  voltage upon reversing the spin current in the GaAs channel.  This sign change is already seen by comparing Figs.~1e and 2c. In Fig.~1 we show data obtained from measurements in which the right Fe electrode is used for spin injection and the left Fe electrode for spin detection (see Fig.~1a). Measurements shown in Fig.~2 were performed with reversed roles of the two Fe electrodes and, therefore, with the reversed orientation of the spin current (see Fig.~2a). As expected, this has no effect on the sign of the measured non-local spin valve voltage while the iSHE voltage changes sign for the two spin-current orientations. In Fig.~3 we provide an additional consistency check of the sign of the iSHE voltage. Here we show measurements in which the sign of the spin current is reversed by using the same Fe electrode for injection but biasing it with the left or right Au contact, respectively. The corresponding experimental setups are shown in Figs.~3a,b and  the data plotted in Fig.~3c confirm that the sign of the measured Hall voltages is opposite for opposite orientations of the spin current in the GaAs channel.

\begin{figure}[h]
\hspace*{0cm}\epsfig{width=0.6\columnwidth,angle=0,file=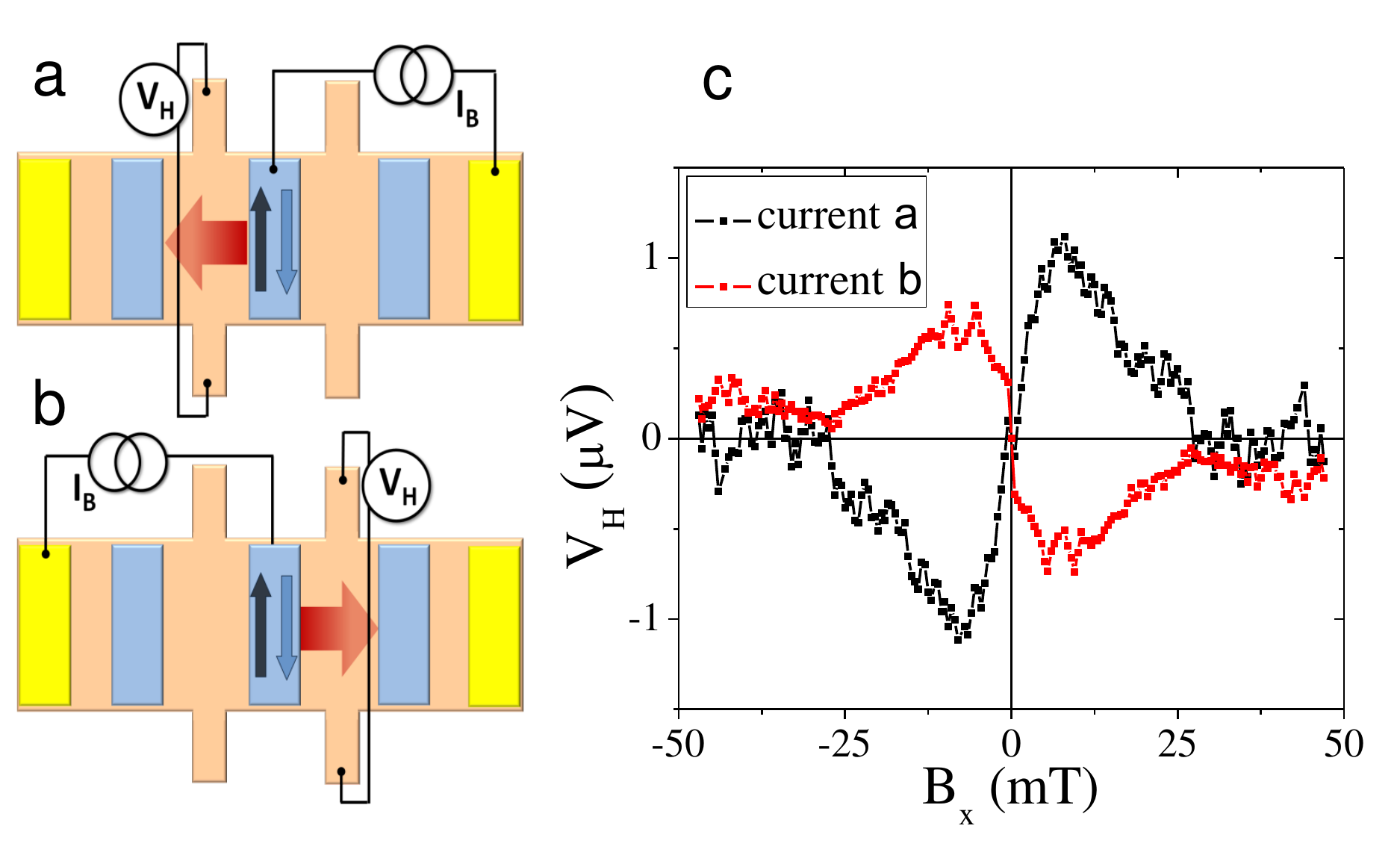}

\vspace*{0.5cm}
\caption{(a),(b) Schematics of the experimental setups with opposite orientations of the spin current in the GaAs channel. (c) iSHE measurements in the in-plane hard-axis field $B_x$ showing opposite  signs for the opposite spin-current orientations. The data were measured at $I_B=300$~$\mu$A and magnetic field sweep rate 3~mT/min.}

\label{eb3}
\end{figure}

In experiments shown in Figs.~1-3, the spins in the GaAs are manipulated by the external magnetic field via the Hanle spin precession. The spin current in these measurements is purely diffusive in the part of the GaAs channel between the injection and detection Fe electrodes. On the other side of the channel with the  bias current $I_B$, both the diffusion and drift are present. In Fig.~4 we show measurements in which we apply a bias between the two Au electrodes causing an additional drift current component $I_D$ on both sides of the injection electrode. The corresponding experimental setup is shown in Fig.~4a. The current $I_B$ driven through the injection Fe electrode is kept at a constant value of 300~$\mu$A  while the additional current $I_D$ is set to 0 and $\pm$100~$\mu$A. The experiments illustrate the possibility to manipulate spins in our GaAs channel electrically via the bias dependent drift. As seen in Figs.~4b,c, both the spin polarization measured underneath the Fe detection electrode and the spin current measured by the iSHE depend on the applied bias between the Au electrodes.

Qualitatively, the experimental data shown in Figs.~4b,c can be explained by a shift of the injected spin polarization profile from the injection electrode in the direction towards the Fe detection electrode in the case of $I_D=+100$~$\mu$A. In the experiment with $I_D=-100$~$\mu$A, the drift acts against diffusion on both sides of the injection electrode which makes the spin polarization profile decay more rapidly as we move away from the injection point.  This explains why the detected spin signals are enhanced for positive $I_D$ and suppressed for negative $I_D$. Our experiments demonstrate a method for modulating the output spin signal by electrical means. Among conventional transistors we can therefore find a loose analogy with the bipolar transistor amplifier. The spin current (spin polarization) detected by the iSHE (non-local spin valve effect) is the spin counterpart of the collector current and the additional drift current $I_D$ is reminiscent of the base current in the bipolar transistor. There is, however, a significant basic physics difference between the charge and spin based device. The latter uses the property that spin is not conserved. By applying the drift to electrons, the non-uniform spin-polarization profile along the channel can be shifted and the corresponding spin current increased or decreased  which causes the electrically controlled modulation of the output signal. Note that our electrical spin manipulation is physically distinct from the electrical bias effect utilized previously in the ballistic Si spin channel.\cite{Huang:2007_a} The experiment in Si relied on the long spin lifetime and was based on electrically controlling the electron transit time through the channel relative to the Hanle precession time in an external magnetic field.

We now proceed to the quantitative theoretical discussion of our spin injection, manipulation, and detection experiments. The spin dynamics  in the GaAs channel can be modeled by the spin drift-diffusion equations. For the applied in-plane hard-axis field $B_x$, the spins precess in the $y-z$ plane and the corresponding Hanle curves are obtained by solving,
\begin{eqnarray}
& &\frac{d {s}_y(x)}{dt}+\frac{d}{dx}\left(-D\frac{ds_y(x)}{dx}+v_d(x) s_y(x)\right)+\frac{s_y(x)}{\tau_s}+g\mu_B B^{eff}_x s_{z}(x)=
\dot{S_0}\delta(x)\nonumber \\
& &\frac{d {s}_z(x)}{dt}+\frac{d}{dx}\left(-D\frac{ds_z(x)}{dx}+v_d(x) s_z(x)\right)+\frac{s_z(x)}{\tau_s}-g\mu_B B^{eff}_x s_{y}(x)=0\,,
\label{DDE}
\end{eqnarray}
where the nuclear Overhauser field is included in the total effective field $B^{eff}_x$, as described in the discussion of Fig.~2. Analogous equations apply for the Hanle curves in the  out-of-plane field $B_z$. In Eqs.~(\ref{DDE}), $D$ is the diffusion constant, $v_d$ is the drift velocity, $\tau_s$ is the spin-dephasing time, $g$ is the Land\'e-factor of electrons in GaAs, and $\mu_B$ is the Bohr magneton. The right-hand side of Eq.~(\ref{DDE}) for the $s_y$ component describes the rate of spins parallel to the Fe magnetic easy-axis ($\hat{y}$-axis) injected from the Fe contact to the GaAs channel at $x=0$. 

In our experiments, the drift velocity can be different on the right and left side of the injection electrode, $v_d(x)=\theta(x)v^R_d-\theta(x)v^L_d$, and is determined by the corresponding currents driven on either side of the injector. For a special case of $v^R_d=v^L_d$, the steady state spin density solving Eq.~(\ref{DDE}) is given by the commonly used expression,\cite{Lou:2007_a}
\begin{equation}
s_{y}(x)=\dot{S_0}\int_0^{\infty}\frac{dt}{\sqrt{4\pi D t}}\exp [(-x-v_d t)^2/4Dt]\exp[-t/\tau_s]\times \cos (g\mu_B B_x t/\hbar)\,; 
\label{constant_v}
\end{equation}
$s_z(x)$ is obtained by replacing cosine by sine in the above expression. Assuming the step-like discontinuity in the drift velocity at the injection point, which corresponds to our experimental geometry, the solution of Eq.~(\ref{DDE}) outside the injection point must have the same functional form as the expression (\ref{constant_v}), up to a normalizing factor. (Outside the injection point, Eq.~(\ref{DDE}) has the same form of a homogeneous differential equation for both the constant or step-like $v_d(x)$.) The origin of the renormalization due to $v_d(x)$ with a sharp step at the injection point is that this form of $v_d(x)$ is equivalent to an additional source/sink term in the drift-diffusion equation at the injection point ($d\theta(x)/dx=\delta(x)$). 
As confirmed by our numerical solution of the drift-diffusion equation, the two normalization factors for the right and left spin densities are obtained by matching the spin densities at the injection point and by requiring the same total integrated spin density as in the case of the constant drift velocity, i.e., $\int_{-\infty}^{\infty}dxs_y(x)=\tau_s \dot{S_0}/[1+(\omega_B \tau_s)^2]$ and $\int_{-\infty}^{\infty}dxs_z(x)=\tau_s \dot{S_0}(\omega_B \tau_s)/[1+(\omega_B \tau_s)^2]$. Note that, the conservation of the integrated spin density  is valid for spatially independent spin-dephasing time and magnetic field  in Eq.~(\ref{DDE}).
  
The drift velocities corresponding to our experiments in Figs.~4b,c are given by, $v^R_d=I_D/enA$ and $v^L_d=(I_D+I_B)/enA$ (see Fig.~4a). Here $e$ is the electron charge, $n$ is the electron density in the GaAs channel, and $A$ is the cross-sectional area of the channel. At the low-temperature used in the measurements, the diffusion constant is given by the expression for a degenerate semiconductor, $D=\mu_en/eg(E_F)$, where $\mu_e$ is the electron mobility and $g(E_F)$ is the density of states at the Fermi level in GaAs conduction band with effective mass $m^\ast=0.067$. The mobility $\mu_e=3.5\times 10^3$~cm$^2$V$^{-1}$s$^{-1}$ and density $n=1.1\times 10^{17}$~cm$^{-3}$, and the corresponding diffusion constant $D=2.9\times10^{-3}$~m$^2$s$^{-1}$ and drift velocities were determined using the ordinary Hall measurements in the GaAs channel. (To extract the coefficients from the ordinary Hall data we considered $A=wt$ where the width of the channel $w=20$~$\mu$m and the effective thickness of the conducting GaAs film $t=270$~nm.)

\begin{figure}[h]
\hspace*{0cm}\epsfig{width=0.6\columnwidth,angle=0,file=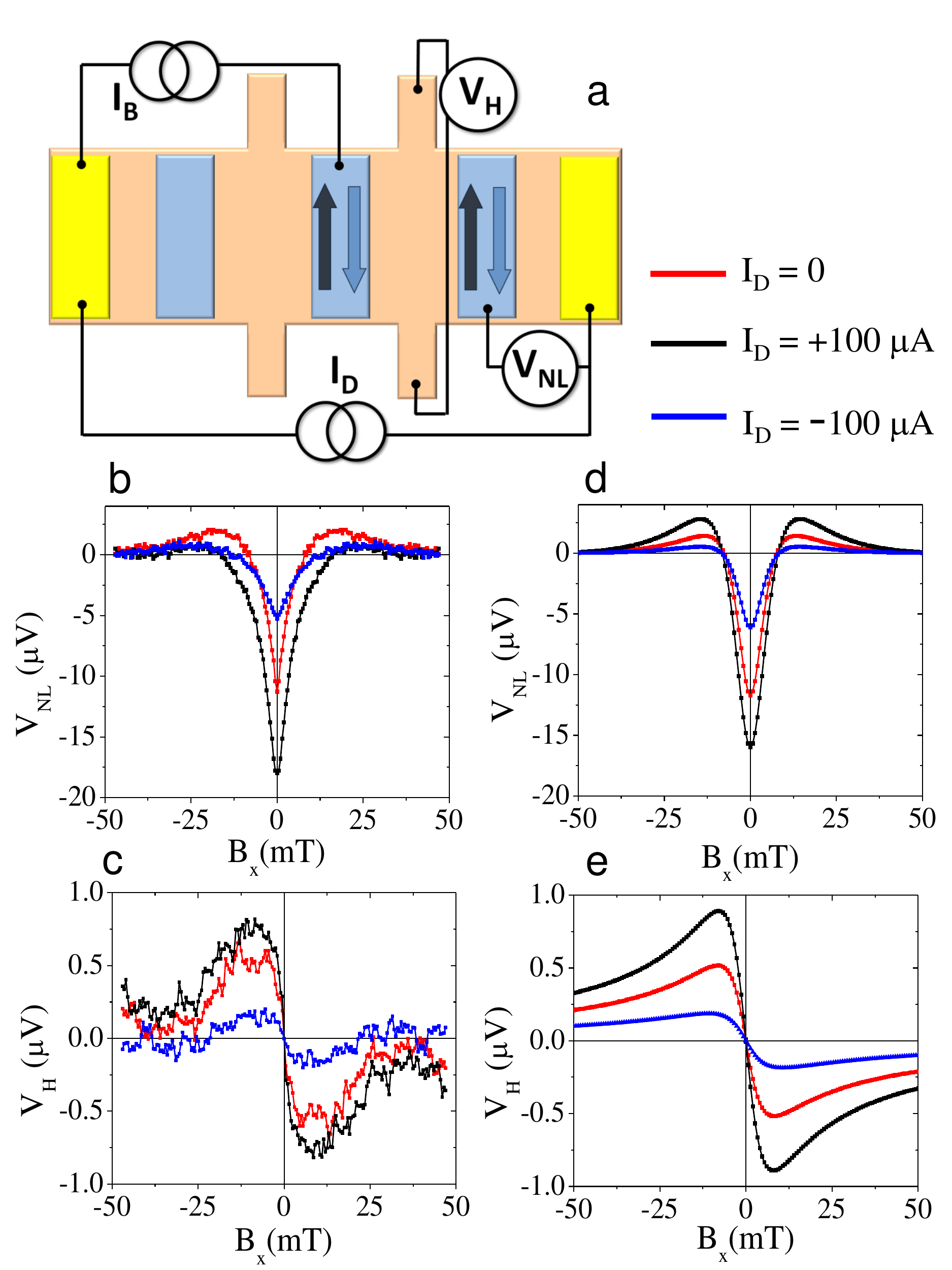}

\vspace*{0.5cm}
\caption{(a) Schematics of the experimental setup. (b),(c) Experimental non-local spin valve and iSHE signals in the in-plane hard-axis field $B_x$ measured at constant spin-injection bias current $I_B=300$~$\mu$A and at three different drift currents $I_D$ depicted in (a). The data were measured at the magnetic field sweep rate 3~mT/min. (d),(e) Theoretical calculations of the non-local spin valve and iSHE signals.
}

\label{eb4}
\end{figure}

The spin-dephasing time $\tau_s=1.65$~ns is obtained by matching the width of the theoretical and experimental Hanle curves. We determined  $\tau_s$ from measurements in the applied out-of-plane hard-axis field $B_z$, i.e., in the geometry where the Overhauser field is negligible. The remaining input parameter needed for obtaining the quantitative values of the theoretical non-local spin valve Hanle curves, shown in Fig.~4d, is the overall normalization factor of the continuous solution of Eq.~(\ref{DDE}) (or equivalently the value of  $\dot{S_0}$). This is obtained by matching the theoretical and experimental spin densities in GaAs underneath the detection electrode. The experimental value is inferred from the difference between the zero field non-local spin valve voltages at parallel and antiparallel magnetization configurations of the injection and detection Fe electrodes considering,\cite{Lou:2007_a}
\begin{equation}
\Delta V_{NL}=2\eta\frac{P_{\rm Fe}P_{\rm GaAs}E_F}{3e}.
\end{equation}
Here $\eta=0.5$ is the spin transmission efficiency of the interface, $P_{\rm Fe} =0.42$ is the polarization of the Fe electrode, and $P_{\rm GaAs}=2s_y(x_d)/n$ is the polarization in GaAs underneath the Fe detection electrode ($x=x_d$). 

The iSHE is proportional to the $\hat{z}$-component of the spin-current given by $j^s_z(x)= -D \vec{\nabla} s_z(x)+v_d(x)s_z(x)$. Since $j^s_z(x)$ depends on the spatial coordinate we have to consider also the response function $F_{cross}(x)$ of the finite-size Hall cross when interpreting the experiments. We performed the numerical evaluation of $F_{cross}(x)$ for our sample geometry using the conformal mapping theory (see Supplementary information).\cite{Thiaville:1997_a,Wunderlich:2001_a} The measured iSHE signal is then proportional to $J^s_z=\int_{-\infty}^{\infty}dxj^s_z(x)F_{cross}(x)/\int_{-\infty}^{\infty}dxF_{cross}(x)$. The spin current and the iSHE voltage are related as, $V_H=ew\alpha J^s_z/\sigma$, where $\alpha$ is the spin Hall angle and $\sigma=ne\mu_e$ is the electrical conductivity of the GaAs channel. The  theoretical $V_H$ plotted in Fig.~4e is obtained by taking $\alpha=1.5\times10^{-3}$ which is a value consistent with the estimates of the skew-scattering Hall angle for the disordered weakly spin-orbit coupled GaAs channel (see Supplementary information). The value is also consistent with electron density dependent spin Hall angles in diffusive GaAs channels reported in optical spin Hall measurements.\cite{Kato:2004_d,Matsuzaka:2009_a} Figs.~4b,d and 4c,e demonstrate the agreement we obtain between the measured and calculated non-local spin valve and iSHE voltages. The theory successfully describes the dependence of the measured spin signals on both the applied magnetic field and on the applied electrical drift current. 

To conclude, we have demonstrated a spin-transistor device based on the iSHE detection of spin currents injected electrically into a semiconductor from a ferromagnetic contact and manipulated by an electric field. Our demonstration was made possible by designing ultrathin-film Fe electrodes with a strong in-plane magnetocrystalline anisotropy to eliminate the ordinary Hall signal and by experimentally introducing an electrical spin manipulation method suitable for diffusive semiconductor channels. We have performed a detailed quantitative analysis of the measured iSHE signals based on complementary non-local spin valve measurements of the injected spin polarization, and on solving the spin drift-diffusion equation and the Hall cross response function.

\section*{Acknowledgment}
We acknowledge support  from EU grants ERC Advanced Grant 268066 - 0MSPIN, FP7-215368 SemiSpinNet, from Czech Republic grants AV0Z10100521, LC510, and Preamium Academiae, and from U.S. grants ONR-N000141110780, NSF-MRSEC
DMR-0820414, NSF-DMR-1105512, and NHARP.

\section*{-- Supplementary information --}
\section{Experimental Methods}

\subsection{Device fabrication and characterization}
\begin{figure}[h]
\centering
\includegraphics[width=0.5\textwidth]{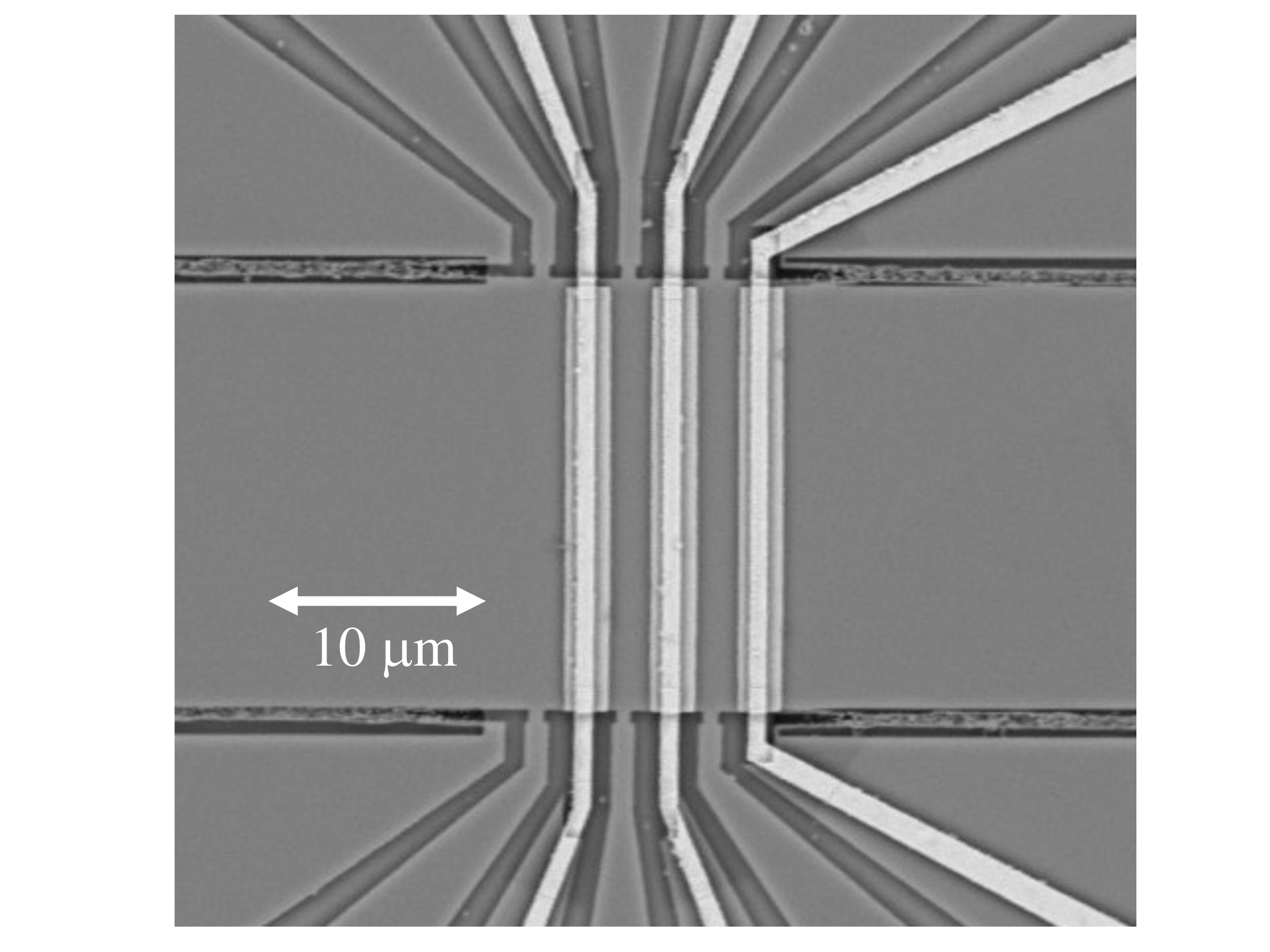}
  \caption{SEM micrograph of the device.}
\label{SEM}
\end{figure}

Fig.~\ref{SEM} shows the SEM micrograph of one of the devices used in this work. Electron-beam lithography and reactive ion etching were used to pattern the lateral GaAs channel with the Hall crosses (grey) and with the magnetic electrodes (white). The injection contacts were first defined by Ti/Au using lift-off. Then the Fe/Al double layer was selectively etched everywhere else. Next the Hall bar was defined using reactive ion etching. Finally the injection contacts were connected with contact pads (outside the figure) using Ti/Au air-bridges. 

\begin{figure}[h]
\includegraphics[width=0.7\textwidth]{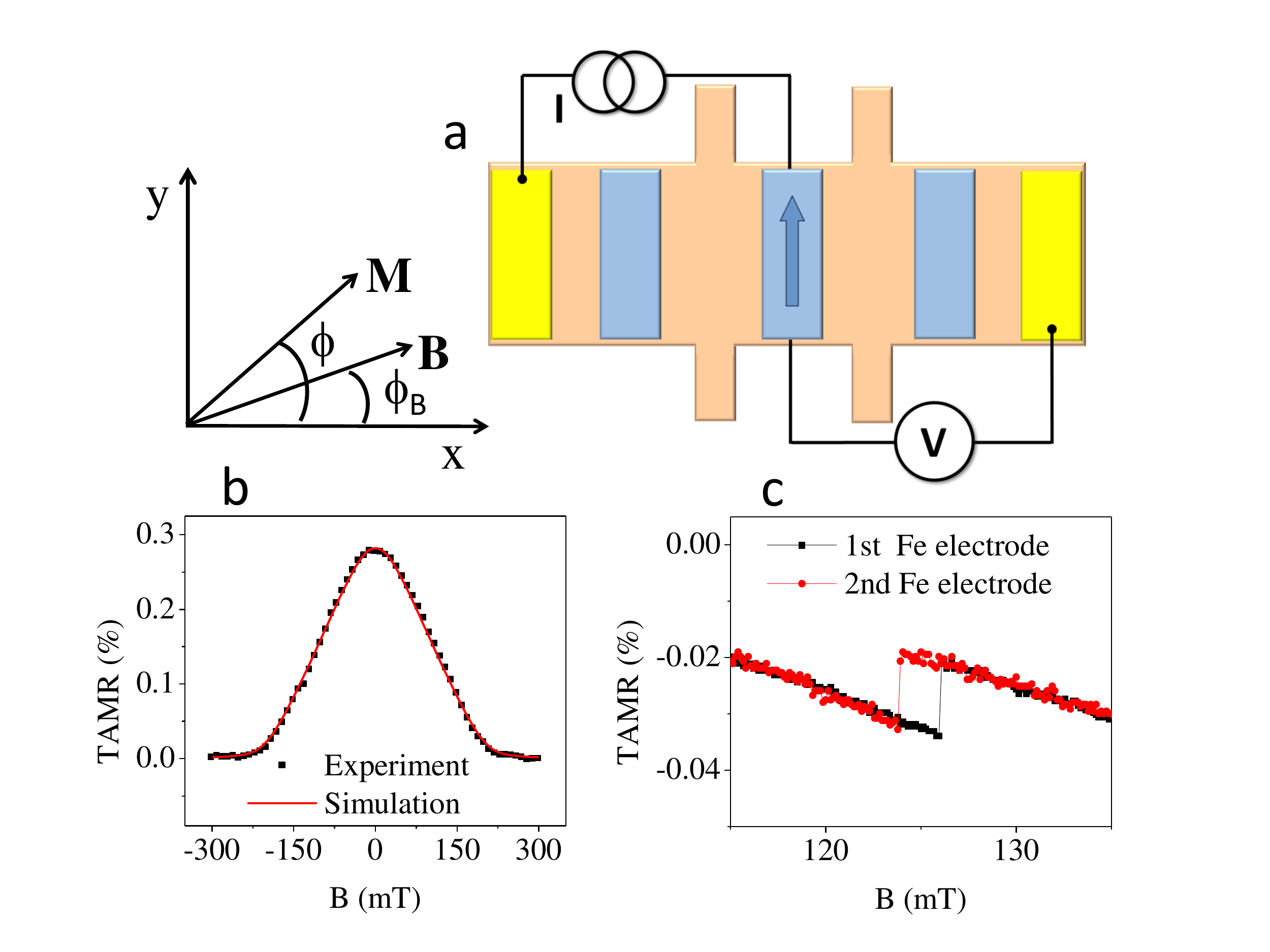}
\caption{(a) TAMR measurement geometry (b) Measured (black squares) and simulated (red curve) TAMR signal of an Fe electrode. (c) Switching of two Fe electrodes monitored by the TAMR.}
\label{supp-TAMR}
\end{figure}

We performed local tunnelling anisotropic magnetoresistance (TAMR) measurements to  monitor the magnetization orientation of the individual Fe electrodes. Fig.~\ref{supp-TAMR}(a) shows the circuit setup we used for the three-point measurements between an individual Fe electrode and the two Au contacts. Current $I$ is sent between the Fe electrode  and one Au contact while voltage $V$ is measured between the Fe electrode and the opposite Au contact. Resistance $R=V/I$ is the resistance of the tunneling contact without the contribution of the channel resistance in this setup.

Fig. \ref{supp-TAMR}(b) shows the resistance variation as a function of the external magnetic field applied closely to the in-plane hard-axis direction. The black squares in Fig. \ref{supp-TAMR}(b) are measurement points and the red curve is the theoretical fitting. The simulation  is based on a single domain model with the total energy $E_{tot}/M_S=B_C/4*\cos^2(2\phi)+B_U/2*\cos^2\phi-B_{ext}\cos(\phi-\phi_B)$.  $M_S$ is the saturation magnetization, $B_C$ and $B_U$ are the cubic and uniaxial anisotropy fields, receptively. $\phi$ and $\phi_B$ are the angles between $x$ (i.e. the$[1\bar10]$ hard-axis) and the in-plane orientation of the magnetization, and the applied magnetic field, respectively. 
Minimizing  $E_{tot}$ for a given $\bold{B}$ determines the position of the magnetization at the particular applied magnetic field. Since the magnetization angle dependence of the TAMR follows\cite{Moser:2006_a} $\sin^2(\phi)$, the resistance can be expressed as $R=R_0+A\sin^2\phi$.  $R_0$ is the tunneling resistance with Fe magnetization along the hard-axis direction, and $A$ is the amplitude of the TAMR signal. The good agreement between experimentally obtained values and the simulation confirms the single domain behavior of the Fe electrodes. The fitting yields the following anisotropy fields: $B_C=20$~mT and $B_U=154$~mT. These values are consistent with previous studies of ultra-thin Fe layers epitaxially grown on GaAs.\cite{Wastlbauer:2005_a} The anisotropy field required to rotate the magnetization from the easy to the hard-axis direction for external field applied along the hard-axis is $B_A=B_U+2B_C=194$~mT. 

An example of detailed TAMR measurements close to the switching fields for two different Fe electrodes are shown in Fig.~\ref{supp-TAMR}(c). The switching is seen as the step of the TAMR whose position and size depends on the applied field angle and the measured electrode. Once the switching fields are known for each electrode, the magnetizations can be switched individually by applying a suitable field. 

\subsection{Experimental extraction of the iSHE}

\begin{figure}[h]
  \centering
    \includegraphics[width=0.6\textwidth]{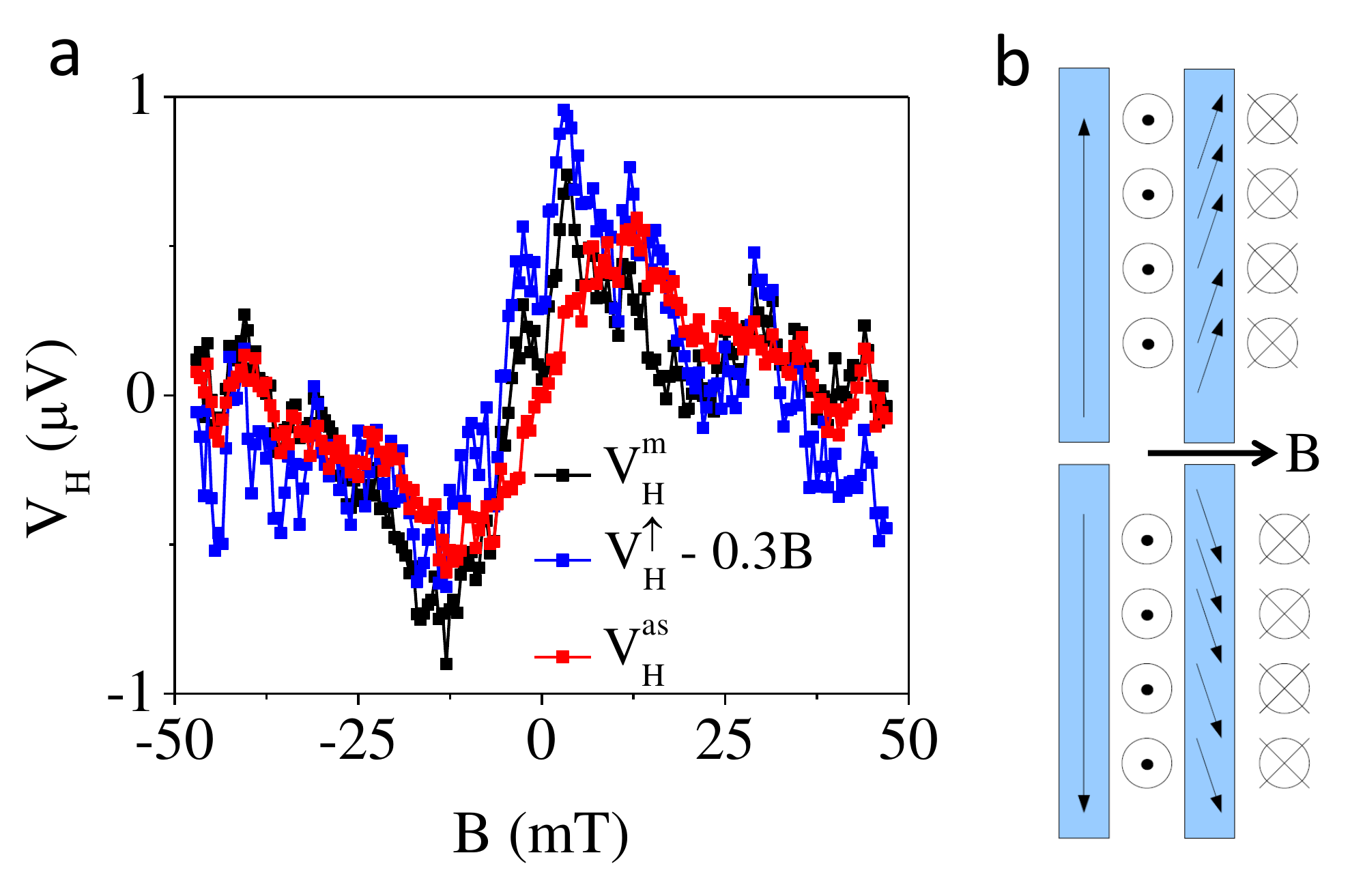}
\caption{
(a) The blue curve shows the measured raw data $V_{H}^{\uparrow}$, obtained by setting the magnetization in the Fe injection electrode in the positive easy-axis orientation, with a subtracted linear  background of 0.3$\mu V$/mT. The black curve shows the pure spin signal $V_{H}^{m}$ obtained by, $V_{H}^{m}=(V_{H}^{\uparrow}-V_{H}^{\downarrow})/2$. The red curve shows the antisymmetrized signal $V_{H}^{as}(B)= [V_{H}^{m}(B)-V_{H}^{m}(-B)]/2$. (b) Schematics of the stray fields when the magnetization is preset along the positive and negative easy-axis orientations and the external hard-axis field is applied.}
    \label{supp-extraction}
\end{figure}

In Fig.~\ref{supp-extraction}(a) we show an example of the extraction of the iSHE signal from the measured raw data. The external magnetic field $\bold{B}$ applied along the in plane magnetic hard-axis is swept from positive to negative values at a rate of 3mT/min. The blue curve shown in the figure are the measured raw data $V_{H}^{\uparrow}$, obtained by setting the magnetization in the Fe injection electrode in the positive easy-axis orientation, with a subtracted linear  background of 0.3$\mu V$/mT. The linear component is attributed to the ordinary (Lorentz force) Hall effect. The ordinary Hall effect may have two contributions in the geometry of our experiment: One due to the vertical part of the trajectory of the electrons passing from the Fe electrode to the GaAs channel, and the other one from the lateral part of the transport in the GaAs channel due to a small out-of-plane misalignment of the sample and the applied magnetic field. (The Hall slope of 0.3$\mu V$/mT corresponds to a small misalignment of few degrees, which is within the experimental error of orienting the sample for the measurement.) The ordinary Hall effect may also have a contribution from the stray field of the Fe electrode. Importantly, both the contributions from the external field and from the stray field  are independent of the Fe magnetization orientation (see Fig.~\ref{supp-extraction}(b)). Therefore, the ordinary Hall effect is eliminated from the measured data by subtracting the measured raw data obtained by setting the magnetizations in the Fe injection electrode in the positive and negative easy-axis orientations. Consistently the resulting signal, $V_{H}^{m}=(V_{H}^{\uparrow}-V_{H}^{\downarrow})/2$, shown as the black curve in Fig.~\ref{supp-extraction}(a) overlaps with the blue curve. The black and blue curves are of pure spin origin and represent the measured iSHE. 

The red curve in Fig.~\ref{supp-extraction}(a) shows the antisymmetric part of the iSHE signal, $V_{H}^{as}(B)= [V_{H}^{m}(B)-V_{H}^{m}(-B)]/2$. The antisymmetrization of the signal allows us to correct for the residual asymmetry of the absolute value of the signal coming from the fact that measurements were done at a rate only approaching the equilibrium nuclear spin polarization. When coming from the large positive fields, the nuclear polarization is higher compared to its equilibrium state, hence it dephases electrons more strongly. When sweeping from zero field to the large negative fields, the opposite situation occurs making the absolute value of the signal weakly asymmetric with respect to positive and negative fields. By comparing the red curve with the blue and black curves we see that the residual asymmetry of the absolute value of the signal is relatively weak and is not obscuring our iSHE data.
	
Note, that the experiments were performed with a small intentional in-plane misalignment ($1-3^{\circ}$) from the in-plane hard-axis which allowed us to preset the magnetization direction without rotating the sample during the measurements. We checked that the obtained iSHE signal was independent of this small in-plane misalignment. Ê

\subsection{Determination of the sign of the spin-polarization}

In order to determine the sign of the injected spin polarization with respect to the magnetization orientation of the injection electrode, we analyzed the response of the injected electron spins to the nuclear spins which were polarized in a controlled way. This technique was previously applied in Ref.~\onlinecite{Salis:2009_a}. The effective magnetic Overhauser field, generated by the dynamically polarized nuclear spins (Eq.~(1) in the main text) is non-zero when the applied magnetic field $\bold{B}$ is not aligned perpendicular to  the electron spin polarization $\langle\bold{S}\rangle$,  i.e. to the magnetization of the injection electrode.  At low applied magnetic fields intentionally misaligned from the hard-axis direction, the Overhauser field can compensate for the applied magnetic field and suppress the resulting spin-decoherence and spin-precession so that a satellite peak appears in the Hanle measurement of the non-local spin valve signal when the Overhauser and external fields cancel each other.\cite{Salis:2009_a} The position of the satellite peak with respect to $B = 0$ determines the sign of the spin-polarization.

\begin{figure}[h]
\centering
\includegraphics[width=0.7\textwidth]{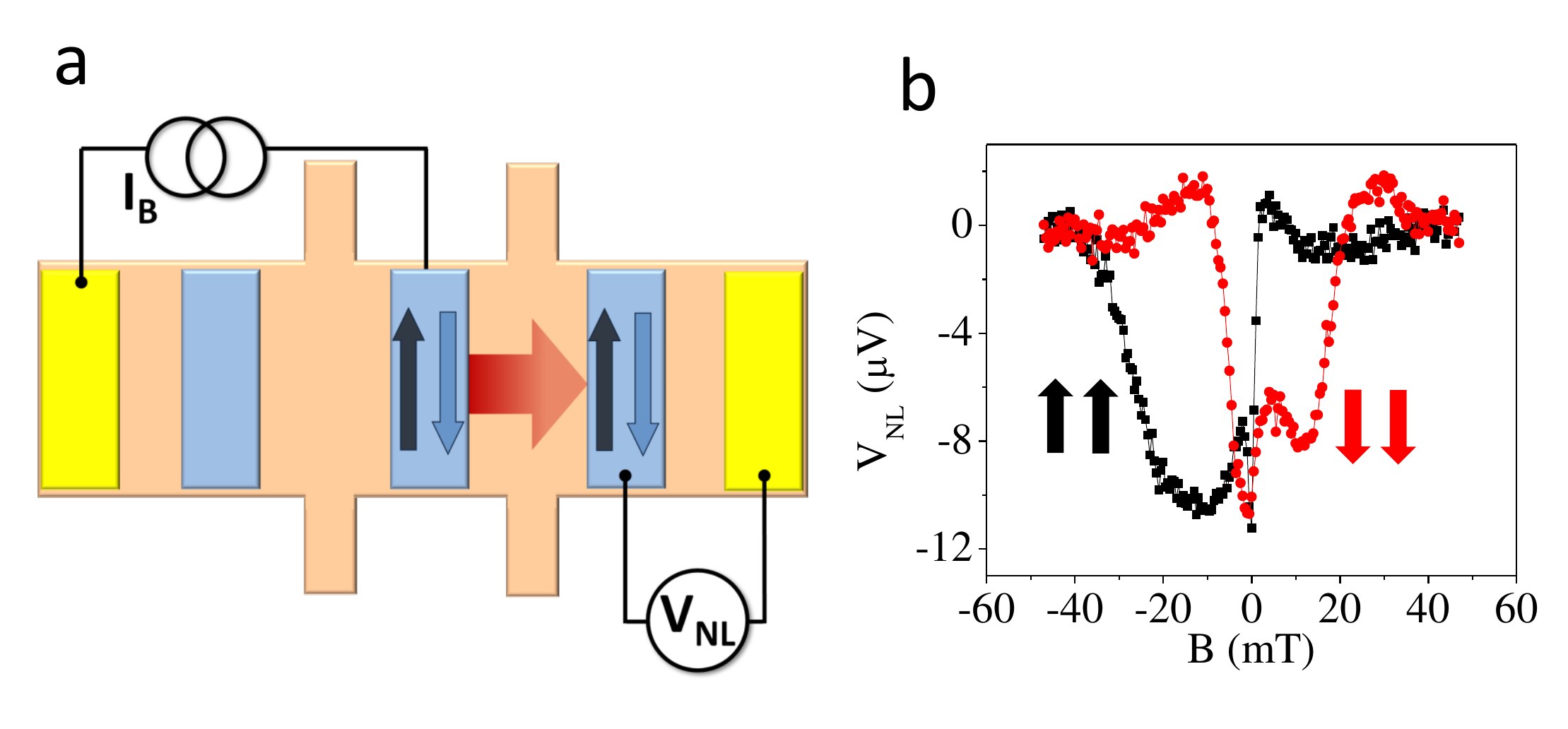}

  \caption{
  (a) Sketch of the nonlocal spin valve Hanle experiment determining the sign of the spin polarization: The applied magnetic field $\bold{B}$ is misaligned with respect to the in-plane magnetic hard-axis by $\sim10^\circ$. 
  (b) Hanle measurements of the non-local spin valve signal for the parallel configuration of injector and detector magnetizations along positive (black) and negative (red) easy-axis orientations. $\bold{B}$  is swept in both cases from +50mT to -50mT at a rate of 3mT/min. 
  }
  
  \label{supp-sign}
\end{figure}

In Fig.~\ref{supp-sign} we show non-local spin valve Hanle signals measured for the parallel configuration of injector and detector magnetizations along positive (black) and negative (red) easy axis orientations. After presetting the magnetizations, a magnetic field misaligned from the in-plane magnetic hard-axis by $\sim10^\circ$  is swept from +50mT to -50mT at a rate of 3mT/min. The Overhauser field is positive  when the magnetization is oriented along  the positive easy-axis orientation because the satellite peak appears at $B < 0$ (black). When the magnetization is along  the negative orientation, the satellite peak is found at $B > 0$ and the corresponding Overhauser field is negative (red). This indicates that the spin polarizations of nuclei and injected electrons are opposite to the magnetization orientation of the Fe spin injection electrode. 

In the case of the external field applied close to the hard-axis direction, the component of $\bold{S}$ in the direction of the external field is given by the tilt of the magnetization towards the applied field.  In the case of the opposite sign of accumulated electron spins to the Fe magnetization, this results in an enhancement of the effective field acting on electrons as introduced in the main text.

\section{Theory Discussion}

\subsection{Spin drift-diffusion equation}

\indent\indent We begin with the drift-diffusion equations in one dimension,
\begin{equation}
\frac{d {S}_i(x,t)}{dt}+\vec{\nabla}\cdot(-D\vec{\nabla}{S_i}(x,t)+\vec{v}_d(x) S_i (x,t))+\frac{S_i(x,t)}{\tau_s}+\gamma\epsilon_{ijk}B_j S_{k}(x,t)=
G_i(x,t)
\label{DDE}
\end{equation}
where $D$ is the diffusion constant, $v_d$ is the drift velocity, $\tau_s$ is the spin-dephasing time, and $\gamma$ is the gyromagnetic ratio. $\vec{G}$ is the injection rate. 
Using these relations to rescale the relevant quantities,
$
x^* = \frac{x}{L_s}$,
$
t^* = \frac{t}{\tau_s}$,
$
\vec{S}^* = \frac{\vec{S}}{G_0\tau_s}$,
$
\vec{G}^* = \frac{\vec{G}}{G_0}$,
$
\vec{B}^* =\gamma\tau_s \vec{B}$,
$
\vec{v_d}^* = \frac{\tau_s}{L_s}\vec{v_d}$, and assuming,
$
L_s^2 = D\tau_S
$ the equation can be reduced to:
\begin{equation}
\frac{d {S^*}_i(x^*,t^*)}{dt^*}+\vec{\nabla}\cdot(-\vec{\nabla}{S^*_i}+\vec{v_d^*} S^*_i )+S^*_i+\epsilon_{ijk}B^*_j S^*_{k}=
\frac{G^*_i}{G_0}
\label{scale_DDE}
\end{equation}

\subsubsection{Zero field case}
The zero magnetic field can be solved analytically in a straight forward way. Following our experimental set-up
 let us assume that $\vec{G}(x)=G_0\delta(x)\hat{z}$, and that there is no magnetic field. Then, in the steady-state case (when the time derivative is zero) we have:
\begin{equation}
-S_y^{*''}(x) + (v^*_d(x)S_y^*(x))' + S_y^*(x)=
\delta(x)
\label{scale_DDE_1D}
\end{equation}
 Here we have written the equation for the $y$-component of S only,  since without magnetic field each component is uncouple, 
 and since we are injecting in the $y$-direction, only $S_y$ is of interest. 
Assuming that $v_d$ is constant,
\begin{equation}
-S_y''(x) + v_dS_y'(x) + S_y(x)=
\delta(x).
\label{scale_DDE_1D_constant_vd}
\end{equation}
Note that $v_d$ refers to the scaled drift velocity as prescribed  before. It is also the only free parameter in this differential equation. 
The solution is
$S(x)=\frac{1}{2\omega}e^{\alpha x}e^{-\omega |x|}$,
where
$
\alpha=\frac{v_d}{2}$ and $
\omega^2=1+\alpha^2$.

We next consider the case that $v_d$ is given by a discontinuous step function, such that
$v_d(x)=v^L_d\theta(-x)+v^R_d\theta(x)$.
The solution is given by:
\begin{equation}
S(x)=\frac{1}{(\omega^R+\omega^L+\alpha^R-\alpha^L)}[e^{(\alpha^L+\omega^L)x}\theta(-x) + e^{(\alpha^R-\omega^R)x}\theta(x)],
\label{solution_scale_DDE_1D_constant_vd}
\end{equation}
\indent\indent where
$
\alpha^L=\frac{v^L_d}{2},\indent\indent\alpha^R=\frac{v^R_d}{2}$,$
(\omega^L)^2=1+(\alpha^L)^2,\indent\indent(\omega^R)^2=1+(\alpha^R)^2$.

\subsubsection{Non-zero field case}

For the case of finite magnetic field it is more straight forward and transparent to proceed in a simple numerical way. For the case of constant drift velocity 
the solution has the form
\begin{equation}
S_{y}(x)=\dot{S_0}\int_0^{\infty}\frac{1}{\sqrt{4\pi D t}}\exp [(-x-v_d t)^2/4Dt]\exp[-t/\tau_s]\times \cos (g\mu_B B_x t/\hbar), 
\label{constant_v}
\end{equation}
with the solution for $S_z$ obtained by replacing cosine by sine in the above expression. 
A sharp step-like form of $v_d(x)$ is equivalent to a source/sink term in the drift-diffusion equation, since 
 a discontinuous drift velocity will tend to accumulate spin  at the discontinuity.
 Within a constant $\tau_s$ approximation, the full steady-state solution of the drift-diffusion equation  can be shown to be  normalized to $\tau_s \dot{S_0}/(1+(\omega_B \tau_s)^2)$ for the y-component and 
 $\tau_s \dot{S_0}(\omega_B \tau_s)/(1+(\omega_B \tau_s)^2)$ for the z-component.
 Hence, the solution for $S_y(x)$ and $S_z(x)$ assuming a step-function behavior of $v_d(x)$ is obtained by Eq.~\ref{constant_v} 
 for constant $v_d$ on the left and the right
 of the drift velocity discontinuity with Eq.~\ref{constant_v} multiplied by appropriate constant factors on the left and on the right to make $S_y(x)$ and
 $S_z(x)$ continuous and normalized correctly.
 
\begin{figure}[h]
\hspace*{0cm}\includegraphics[width=0.5\columnwidth]{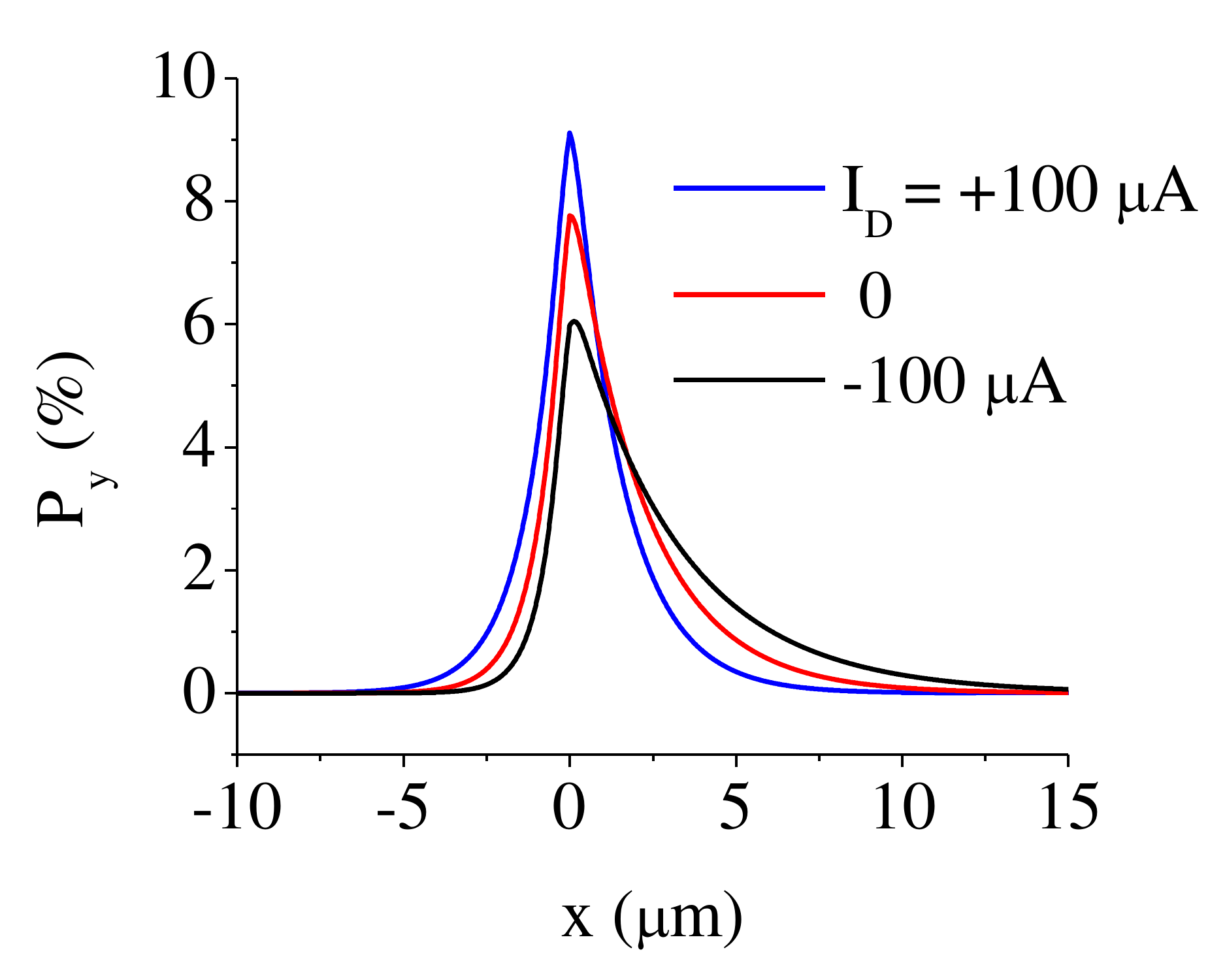}
\caption{Spin polarization profile $P_y(x)=S_y(x)/n$ obtained
by solving the drift-diffusion equations for the experimental parameters of Fig.~4 in the main text.}
\label{figs5}
\end{figure}

 In Fig.~\ref{figs5} we illustrate the $S_y(x)$  for the situation where the additional drift current ($I_D$ in Fig. 4 in the main text) is
 $+100$ $ \mu{\rm A}$, $0$, $-100~\mu{\rm A}$ and the bias current through the injection electrode $I_B=300$ $\mu{\rm A}$, as done in the experiment.
The spin-current generated by this spin accumulation profile is given by 
\begin{equation}
j_i\equiv -D \vec{\nabla} s_i(\vec{r})+v_d(\vec{r})s_i(\vec{r}).
\label{ji}
\end{equation}

\subsection{Hall effect}
The conduction electrons can be modeled through the following effective Hamiltonian:
\begin{equation}
H=\frac{\hbar^2 k^2}{2m}+V_{\rm dis}(\rr)
+\lambda^* {\boldsymbol \sigma} \cdot(\kk\times{\bf \nabla}V_{\rm dis}(\rr)),
\label{Ham}
\end{equation}
where the $m=0.067 m_e$
and $V_{\rm dis}$ is the disorder potential modeled by uncorrelated delta scatterers of strength $V_0$ and density $n_i$.
 For GaAs $\lambda^*=5.3\,$ \AA$^2$.\cite{Knap:1996_a,Winkler:2003_a}
The Hall effect signal can be understood within the theory of the  anomalous Hall effect (AHE). The contributions to the AHE in SO coupled
systems with non-zero polarization can be classified in two types: the first type arises from the SO coupled quasiparticles interacting with
the spin-independent disorder and the electric field, and the second type arises from the non-SO coupled part of the quasiparticles scattering
from the SO coupled disorder potential. The contributions of the first type do not dominate the physics of
this weakly spin-orbit coupled system, and is therefore not included in Eq. \ref{Ham}.

The contributions of the second type, i.e. from interactions with the SO
coupled part of the  disorder,\cite{Nozieres:1973_a,Crepieux:2001_b} are due to the
anisotropic scattering, the so called extrinsic skew-scattering,
and is obtained within the second Born approximation treatment of the collision
integral in the semiclassical linear transport theory:\cite{Nozieres:1973_a,Crepieux:2001_b}
\begin{equation}
|\sigma_{\rm xy}|^{\rm skew}=\frac{2\pi e^2\lambda^*}{\hbar^2}V_0 \tau n^2.
\end{equation}
Using the relation for the mobility $\mu=e\tau/m$ and the relation between $n_i$, $V_0$,  and $\tau$, $\hbar/\tau=n_iV_0^2m/\hbar^2$,
the extrinsic skew-scattering contribution to the iSHE angle due to a pure spin-current, $\alpha\equiv \rho_{xy}/\rho_{xx}\approx\sigma_{xy}/\sigma_{xx}$,  can be written as
\begin{eqnarray}
\alpha^{\rm skew}&=&2.44\times 10^{-4}
\frac{\lambda^*[{\rm \AA^2}]n_{\rm 2D}[10^{11}{ \rm cm}^{-2}]}{\sqrt{\mu[10^3{ \rm cm}^2/{\rm Vs}] n_{\rm 2D-i}[10^{11}{ \rm cm}^{-2}]}}\nonumber\\
&\sim& 3.8 \times 10^{-3},
\end{eqnarray}
where we have used $n_{2D}=n t=3.0\times 10^{12} {\rm cm}^{-2}$, $t=270$ nm is the GaAs film thickness, $n=1.1\times 10^{17} {\rm cm}^{-3}$,
$\mu=3.5 \times 10^3 {\rm cm}^2/{\rm Vs}$, and $n_{2D-i}\approx 3.0\times10^{12} {\rm cm}^{-2}$.

\subsection{Hall Response Function}
When dealing with non-uniform currents, it is non-trivial to relate the measured Hall voltage signal with the Hall angle or Hall coefficient of the system.\cite{Wunderlich:2001_a,Thiaville:1997_a} In geometries where the Hall probe width and the channel width are of similar magnitude the current density near the the cross can contribute to the Hall signal more significantly and one must solve the full equations relating the current density and the fields. For the case where anomalous Hall effect is considered in addition to the normal Hall effect we can write
\begin{equation}
\vec{j}(x,y)=\rho^{-1}(-\vec{\nabla}V(x,y)+\vec{j}(x,y)\times[R_0\vec{B}(x,y)+4\pi R_s \vec{M}(x,y)]\,\,,
\end{equation}
where $\rho$ is the diagonal electrical resisitvity of the layer, $\vec{B}$ and $\vec{M}$ are the local magnetic induction and magnetization, and
$R_0$ and $R_s$ are the normal and anomalous Hall coefficients. Here we have assumed a thin film geometry such that $j_z=0$ and the problem is reduced to two dimensions. 

We then must solve the Maxwell equations for a static magnetic field
\begin{equation}
\nabla^2 V=0\,\,\,{\rm and}\,\,\,\vec{\nabla}\cdot\vec{j}=0\,\, ,
\end{equation}
with the boundary condition that the current is zero at the insulating cross boundaries, $i.e.$ $\vec{j}\cdot\hat{n}=0$ at the boundaries. 

We follow here a similar procedure as in Ref.~\onlinecite{Thiaville:1997_a}. Taking $\beta=\rho^{-1}[R_0B(x,y)+4\pi R_s {M}_\perp(x,y)]$ the above 
equations reduce to 
\begin{equation}
(1+\beta^2))\nabla^2 V+(1-\beta^2)\left(\frac{\partial \beta}{\partial x}\frac{\partial V}{\partial y}-\frac{\partial \beta}{\partial y}\frac{\partial V}{\partial x}\right)
-2\beta \left(\frac{\partial \beta}{\partial x}\frac{\partial V}{\partial x}+\frac{\partial \beta}{\partial y}\frac{\partial V}{\partial y}\right)=0
\end{equation}
and the boundary condition to $\nabla_\perp V=-\beta\nabla_{||} V$. Since $\beta$ tends to be small in most systems of interest we can treat the problem 
perturbatively, $V=V_0+V_1+\dots$, whose first two components solve
\begin{equation}
\nabla^2 V_0=0,
\end{equation}
with $\partial V_0/\partial n=0$ at the boundaries and
\begin{equation}
\nabla V_1=\frac{\partial \beta}{\partial y}\frac{\partial V_0}{\partial x}-\frac{\partial \beta}{\partial x}\frac{\partial V_0}{\partial y},
\end{equation}
with $\nabla_\perp V_1 =-\beta \nabla_{II} V_0$ at the boundaries. Solving these equations for the case of a delta-like magnetic field at position $(x,y)$ yields the Hall response function which  
can then be convoluted with the non-constant magnetic field or magnetization to obtain the total Hall signal expected.

In our case, we are considering the response to a pure spin-current which can be considered as two fully spin polarized charge
currents with opposite polarities and direction. This allows us to use the result shown here, only ignoring the small fraction contributing from 
the polarized charge current on the left of the injection point far away, relative to the spin-diffusion length, from the detecting Hall bar.

\subsubsection{Solution of $V_0$}
The solution of $V_0$ for the Hall cross bar geometry can be done using the conformal mapping technique. In here one uses the
fact that for any analytical function $f(z)=u(x,y)+iv(x,y)$ in the complex z-plane, the conjugate functions $u$ and $v$ solve the Laplace 
equation. Then the problem reduces to finding the analytical function that solves the boundary conditions of the problem. To do so
one can do a conformal mapping of the region of interest in the z-plane to a much easier configuration in another complex plane, e.g. parallel plate. The conformal mapping preserves the boundary conditions and the solution in the z-plane can be obtained by mapping
backwards the trivial solution in the complex plane.

For the case of the Hall cross, or any polygon structure for that matter, one uses the Swartz-Christoffen transformation. We impose the boundary condition for $f(z)=u(x,y)+iv(x,y)$ such that $v$ is equal to $\pm \pi$ at the boundary edges along the channel. 
The Swartz-Christoffen transformation transforms any interior region of a polygon 
(even ones with open boundaries where the vertex is at infinity) onto the upper half of the complex plane with the vortices mapped to points in the real axis. For the present configuration the map reads
\begin{equation}
\frac{d z}{d w}= C i \frac{\sqrt{z^2-a^2}}{z^2-b^2},
\end{equation}
where $a$, $b$, $C$, are constants adjusted such that the the vortices map correctly to the right places. This maps the problem to a system where the potentail is $\pm \pi$ on the real axis which can the be mapped simply via a second transformation 
\begin{equation}
\xi=\log\left(\frac{w-1}{w+1}\right )-i\pi,
\end{equation}
which is simply the solution of a parallel plate capacitor. Depending on the geometry some fraction of the current avoids the central region of the cross bar. 

\begin{figure}[h!]
\hspace*{0cm}\includegraphics[width=0.65\columnwidth]{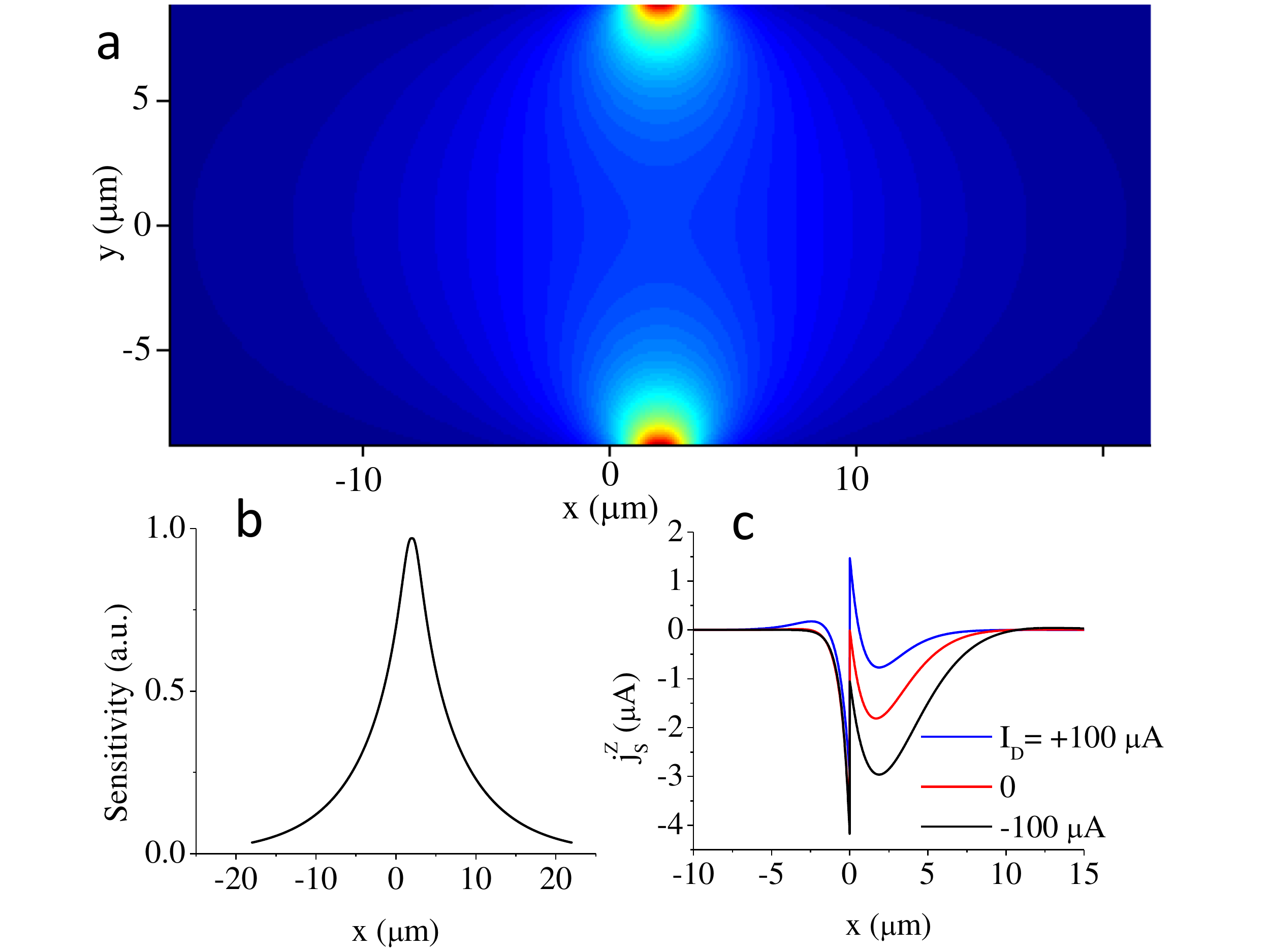}
\caption{(a) Hall cross response function in the $x-y$ plane for the geometry of the measured device. (b) Hall cross response function averaged over the channel width ($y$-direction). (c) Spin-current profile $j_z^s(x)$ obtained
by solving the drift-diffusion equations for the experimental parameters of Fig.~4 in the main text.}
\label{figs6}
\end{figure}

\subsubsection{Solution of the Hall response function: $V_1$ for a Dirac delta function magnetic field}
Since the equations that we are solving, to first order, retain the principle of superposition we can solve for the Hall response function due
to a delta function magnetic field.

The Hall response function is found (see App. in Ref.~\onlinecite{Thiaville:1997_a}) to be
\begin{equation}
F_{cross}(x,y)=\frac{\pi}{2}\frac{{\rm Im}\sqrt{w^2-b^2}}{| w^2-a^2|}.
\end{equation}
The above response function is normalized to $2\pi$. 
The response function for our experimental set-up is shown in Fig.~\ref{figs6}(a). In realistic situations there is no current with about $100\,{\rm nm}$ of the edge
and we therefore exclude the sharp part of the response function near these edges. Since we are only considering the spin accumulation in one dimension, we
can average this response function in the y-direction, as shown in Fig.~\ref{figs6}(b).
Solving for the spin-current from the drift-diffusion equations in one dimension, shown in Fig.~\ref{figs6}(c),
using Eq.~\ref{ji}, we then convolute this result with the response function integrated along the $y$-direction to obtain the measured spin-current
$J_z^s=\int_{-\infty}^{\infty} dx j_z^s(x) F_{cross}(x)/\int_{-\infty}^{\infty} dx  F_{cross}(x)$ which is related to the iSHE voltage
by $V_H=e\omega\alpha J_z^2/\sigma$ as described in the main text.

\end{document}